\documentclass[aps,prl,twocolumn,superscriptaddress]{revtex4-2}
\usepackage{amsmath,amsfonts,amssymb,bm,graphicx,subfigure}

\usepackage[notrig]{physics}
\usepackage{CJKutf8}
\usepackage{orcidlink}
\usepackage{hyperref}
\usepackage[final]{changes}

\UseRawInputEncoding       

\newcommand{\ud}{\mathrm{d}}

\usepackage{color,soul}

\begin{document}

\begin{CJK*}{UTF8}{gbsn} % Use default fonts from CJK (see below)

\title{Transfer of active motion from medium to probe\\ via the induced friction and noise}

\author{Ji-Hui Pei (裴继辉) \orcidlink{0000-0002-3466-4791}}
\affiliation{Department of Physics and Astronomy, KU Leuven, 3000, Belgium}
\affiliation{School of Physics, Peking University, Beijing, 100871, China}
\author{Christian Maes \orcidlink{0000-0002-0188-697X}}
\email{christian.maes@kuleuven.be}
\affiliation{Department of Physics and Astronomy, KU Leuven, 3000, Belgium}

%\date{\today}

\begin{abstract}
Can activity be transmitted from smaller to larger scales? We report on such a transfer from a homogeneous active medium to a Newtonian spherical probe. The active medium consists of faster and dilute self-propelled particles, modeled as run-and-tumble particles in 1D or as active Brownian particles in 2D. We derive the reduced fluctuating dynamics of the probe, valid for arbitrary probe velocity, characterized by velocity-dependent friction and noise. 
In addition to a standard passive regime, we identify peculiar active regimes where the probe becomes self-propelled with high persistence, and its velocity distribution begets peaks at nonzero values. These features are quantitatively confirmed by numerical simulations of the composite probe-medium system.
The emergence of active regimes depends not only on the far-from-equilibrium nature of the medium but also on the probe-medium coupling.
Our findings reveal how, solely via the induced friction and noise, persistence can cross different scales to transfer active motion.
\end{abstract}
\maketitle
\end{CJK*}

\textit{Introduction.}
\deleted{Bridging different levels of physical description is a hallmark of statistical mechanics. 
Understanding how properties of microscopic evolutions, when combined with statistical considerations, penetrate the realm of mesoscopic and macroscopic physics and give rise to new emerging phenomena is one of the biggest challenges.  A specific and important paradigm is the
Brownian motion of a micron-sized colloidal particle suspended in a fluid at rest, where the environment is passive and the fluctuating motion is purely thermal. In this work, we explore a similar setup but with an active environment, uncovering very different reduced dynamics.
}
Active systems \cite{Toner2005,Ramaswamy2010,Marchetti2013,Bechinger2016,Gompper2020,Vrugt2025,Granek2024} drive themselves far from equilibrium by  local energy consuming processes \cite{Gompper2020}.
To investigate how active systems influence phenomena at \replaced{larger scales}{a larger scale of length, mass or time}, 
studying a heavy probe immersed in an active medium (or active bath) has attracted a lot of attention \cite{Granek2024,Wu2000,Chen2007,Leptos2009,Rafai2010,Kurtuldu2011,Valeriani2011,Mino2011,Mino2013,Kaiser2014,Maggi2014,Patteson2016,Argun2016,Kurihara2017,Maggi2017, Jerez2017,Katuri2021,Paul2022,Boriskovsky2024,Caprini2024,Gregoire2001,Loi2008,Underhill2008,Angelani2010,Foffano2012,Mallory2014,Suma2016,Knezevic2020,Ye2020,Shea2022,Dor2022,Feng2023,Jayaram2023,Zakine2017,Burkholder2017,Burkholder2019,Kanazawa2020,Reichert2021,Abbaspour2021,Peng2022,Ion2023,Dhar2024,Dolai2024,Zhao2024,Sarkar2024,act,Steffenoni2016,Granek2022,Solon2022}. 

A scalar active medium can be modeled as a collection of self-propelled particles that show persistence in velocity, such as run-and-tumble and active Brownian particles \cite{Erdmann2000,Tailleur2008,Romanczuk2012,Solon2015,Malakar2018,Basu2018,Demaerel2018,Caprini2022}.
While we know how an equilibrium bath leads to a passive probe motion, we enquire here
whether and how a probe can inherit active motion from an active bath, that is, endowed with a persistent velocity. 
To fully understand this question requires knowledge of the reduced dynamics for the probe after theoretically integrating out the active bath.

The problem of integrating out the motion of active particles
can proceed via several methods \cite{D’Alessio2016,act,Solon2022,Ion2023}. 
For an active bath consisting of effectively independent active particles, existing results \cite{Steffenoni2016,Solon2022,Tanogami2022,Pei2025,Granek2022} suggest the following (or similar) underdamped Brownian dynamics for a heavy spherical probe with mass $M$ and velocity $\mathbf v$, 
\begin{equation}\label{sd}
    M\dot{\mathbf v} = -\gamma\mathbf v + \sqrt{2B}\,\boldsymbol\xi ,
\end{equation}
with linear friction coefficient $\gamma$, standard white noise $\boldsymbol\xi$ and its intensity $B$. 
Such dynamics with linear friction are valid only for small probe speeds. 
For an active medium, $\gamma$ and $B$ do not satisfy the standard Einstein relation \cite{Harada2006,Maes2014}, and it has recently been found that the linear friction coefficient $\gamma$ can be negative \cite{Granek2022,Kim2024,Granek2024,Pei2025}, indicating that the probe would accelerate when starting from a small velocity. 
After acceleration to higher speed, friction can no longer be described in linear form $-\gamma \mathbf v$. 
Therefore, when $\gamma<0$,  the Brownian dynamics \eqref{sd} fails.
Obtaining the correct reduced fluctuating motion of a spherical probe in a scalar active medium has remained a significant unsolved problem.

In this Letter, we identify the cause of the breakdown of \eqref{sd}
and derive the corrected reduced dynamics\replaced{, which includes}{ through a frame transformation prior to a quasistatic expansion. 
This procedure yields} three components: a first-order nonlinear velocity-dependent friction (also reported in \cite{Kim2024}), its second-order correction, and a velocity-dependent noise.
\replaced{We classify the probe motion into a passive regime and}{From the corrected reduced dynamics, we observe both a passive regime and the emergence of} notable active regimes.
In the active regimes, activity is transferred to the probe: a 1D probe follows run-and-tumble motion or run-and-stop motion, while a 2D probe either exhibits active Brownian motion or switches randomly between active and passive Brownian motions.
\deleted{These results for the reduced dynamics are quantitatively confirmed by simulations. }

This activity transmission hinges on the probe-medium coupling and the dimension. 
Unlike 1D where a soft coupling is necessary, \deleted{we find that }realistic hardcore (Lennard-Jones) interactions suffice to transfer activity in 2D, 
and we provide an experimental proposal.

\textit{Setup and general structure.}
The active medium is spatially homogeneous and\deleted{ considered in a} dilute\deleted{ limit}, consisting of $N$ independent overdamped self-propelled particles with positions $\mathbf z^a$ ($ a=1,\dots,N$ labels different particles). 
We take a periodic boundary on $(-L/2,L/2]^d$. 
The infinite size corresponds to the limit of $L,N\rightarrow\infty$ with fixed low density $n=N/L^d$.
The Newtonian underdamped probe has position $\mathbf q$ and velocity $\mathbf v $. 
The interaction force between medium particles and probe derives from an isotropic potential, $F(\abs{\mathbf q-\mathbf z})=-U^\prime(\abs{\mathbf q-\mathbf z})$. The equations of motion are, 
\begin{equation}\label{or}
    \begin{aligned}
    \mu\dot {\mathbf z}^a &= F(r^a)\hat{\mathbf r}^a +\mathbf A^a, \\
    M \dot {\mathbf v} &= -\sum_{a=1}^N F(r^a)\hat{\mathbf r}^a , \quad \dot {\mathbf q} = \mathbf v
    \end{aligned}
\end{equation}
where $\mathbf A^a$ represents independent (for now unspecified) self-propulsion forces on the individual active particles; 
$\mathbf r^a=\mathbf z^a-\mathbf q$ is the relative position between the $a$-th medium particle and the probe, and $\hat {\mathbf{r}}^a=\mathbf r^a/r^a$ denotes its direction.
$\mu$ is the inverse mobility of medium particles.  
The probe mass $M$ is assumed large. We aim at a reduced description for the probe.

Existing studies \cite{Steffenoni2016,Solon2022,Tanogami2022,Pei2025} obtain the Brownian dynamics \eqref{sd} by assuming a time-scale separation between the probe and the medium particles. 
Yet, that assumption is not always valid because position $\mathbf q$ may not change slowly; that is, $\mathbf v=\dot {\mathbf{q}}$ may not be small (although the velocity $\mathbf v$ itself is usually slow since $\dot{\mathbf{v}}$ is small for a heavy probe).  
%Especially, when the linear friction coefficient $\gamma<0$ (possible for an active medium \cite{Granek2022,Pei2025}), the probe gradually accelerates to a high velocity, destroying the time-scale separation and the general validity of Eq.~\eqref{sd}.

% Actually, even for equilibrium baths, there are special cases where the probe position is not a slow variable, e.g., when the probe is externally driven to a high velocity. 
% In \cite{Itami2017}, systematic theoretical analysis (on equilibrium baths) has shown that velocity-dependent friction and noise appear when the position is not slow.
% However, to obtain explicit expressions for velocity-dependent friction and noise is generally challenging, let alone for an active medium. 

We address this problem by the following \deleted{physical }procedure. 
%\added{Even in equilibrium, there are cases that only the velocity is slow, e.g., relaxation of an initially fast probe \cite{Itami2017}.}
For a homogeneous medium, 
$\mathbf q$ can be eliminated by the change of variables  ${\mathbf z}^a\rightarrow {\mathbf r}^a={\mathbf z}^a-{\mathbf q}$, which shifts the medium motion to the moving reference frame where the probe remains at the origin. 
The equations of motion \eqref{or} now become 
\begin{equation}\label{medium_dynamics}
\begin{aligned}
        \mu\dot{ \mathbf r}^a &=-\mu \mathbf v +F(r^a)\hat{\mathbf r}^a  +\mathbf A^a, \\
        M\dot{ \mathbf v} &= -\sum_{a=1}^N F(r^a)\hat{\mathbf r}^a .
    \end{aligned}
\end{equation}
The probe position $\mathbf q$ does not appear anymore. 
For a heavy probe, \deleted{unlike position $\mathbf q$, }the velocity $\mathbf v$ is always slow, 
and the time scale of $\mathbf v$ ($\tau_v$) is much larger than that of the $\mathbf r^a$ ($\tau_r$), characterized by a small constant $\epsilon=\tau_r/\tau_v$; see supplemental material (SM) \cite{SM}.

We can thus start from \eqref{medium_dynamics} to safely integrate out the medium particles as in the usual quasistatic approximation.  As explained in SM \cite{SM}, up to $O(\epsilon^2)$, we obtain a nonlinear fluctuating dynamics \cite{Itami2017} of the probe, where velocity-dependent friction and noise appear\deleted{, all expressed as explicit correlation functions}.  
\deleted{This powerful reduced dynamics indicates that the active motion can be transmitted from the medium solely via the effects of friction and noise.  
It makes the theoretical treatment of modeling active particles by velocity-dependent friction and noise come true \cite{Erdmann2000,Romanczuk2012}.}
In what follows, we \replaced{present our results}{focus on the reduced dynamics} in 1D and 2D for specific active media.

\textit{1D run-and-tumble medium.}
In 1D, we consider an active medium consisting of run-and-tumble particles \cite{Solon2015,Malakar2018,Demaerel2018,Dhar2019}. 
The equation of motion \eqref{medium_dynamics} of one medium particle in the moving frame becomes
\begin{equation}\label{1dm}
   \mu\dot r = F(r)+\mu u \sigma -\mu v,
\end{equation}
where $\sigma =\pm 1$ flips randomly at a Poisson rate $\alpha$, and it indicates the direction of the constant self-propulsion speed $u > 0$. 
After integrating out medium particles, the reduced dynamics of the probe is given by a nonlinear equation \cite{SM},
\begin{equation}\label{1dr}
\begin{aligned}
    M\dot{{v}}(t) = &-f ({v}(t)) +\sqrt{2B({v}(t))}\,\xi(t) \\
    &-\frac 1M G({v}(t)) + \frac{1-\eta}{M} B^\prime({v}(t)) ,
    \end{aligned}
\end{equation}
where $\xi(t)$ is a standard white noise; $\eta$ depends on the discretization convention of the stochastic differential equation: $\eta=0$ for It\^o, $\eta=1/2$ for Stratonovich, and $\eta=1$ for anti-It\^o. The last term originates from the multiplicative noise.
The (first-order) nonlinear friction $f(v)$, the noise intensity $B(v)$, and the second-order correction of the friction $G(v)$
are given\deleted{ explicitly} in the End Matter.
They are\deleted{ all} expressed as expectation values in the ``fixed-$v$" dynamics (Eq. \eqref{1dm} with fixed $v$) of a single medium particle.
The term $f(v)$ is of order $\epsilon$, and its linear part $\gamma=f^\prime(v)\eval_{v=0}$ recovers the friction coefficient in previous studies \cite{SM}. In that sense, we call $f$ the nonlinear friction. 
\deleted{On the other hand, }$G(v)$ and $B(v)$ are of order $\epsilon^2$.
The total nonlinear friction up to $O(\epsilon^2)$ is then $g(v)=f(v)+G(v)/M$, where $G(v)/M$ can be ignored in a qualitative approach but not in a quantitative analysis. 
The presence of velocity-dependent friction and noise is the main characteristic of the above dynamics, making it valid for arbitrary $v$.

\replaced{The nonlinear friction $f(v)$ is not always positive}{The positivity of the nonlinear friction $f(v)$ is not ensured}, which may cause the active motion of the probe.
In the End Matter, we provide an intuitive analysis of $f(v)$.
There are two requirements for $f(v)$ to exhibit \replaced{negativity}{negative values}: the persistence length of medium particles should be large compared to the interaction range, $u/\alpha\gg R$; the coupling should not be hardcore, so that medium particles can pass through the probe.

From expressions of $f$, $B$, and $G$, one can straightforwardly calculate them numerically. 
For the calculation, we choose the probe-medium coupling to be a soft repulsive interaction $F(r)=k\sin (\pi r/R)$ for $\abs{r}<R$ and $F(r)=0$ for $\abs{r}>R$, with range $R$ and strength $k$. 
The plots of $f(v)$ and $B(v)$ are shown in Figs.~\ref{fig:1d}(a)-(b). 
The singular turn over in Fig.~\ref{fig:1d}(b) is explained in the End Matter.
In SM \cite{SM}, we supplement more figures,\replaced{ showing}{ which help to show} how the negative region in $f (v)$ depends on the medium persistence and the interaction strength. 

\begin{figure*}[!htp]
   \centering
   \includegraphics[scale=0.071]{pic1d.pdf}
   \caption{For 1D run-and-tumble medium with interaction $F(r)=k\sin(\pi r/R)$ within range $r<R$: (a) Friction $f(v)$ and noise intensity $B(v)$ (per medium particle) with different flip rates $\alpha$, showing regimes (R1) and (R2a). Parameters are $L=10$, $R=0.5$, $k=2.4$, $u=3$, $\mu=1$.
   (b) Same for $k=3.3$, corresponding to regimes (R1) and (R2b).
   (c) Stationary velocity distribution $\rho^\text{st}(v)$ of the probe, from the simulation (blue), from the nonlinear dynamics \eqref{1dr} (red dashed), and from the effective active dynamics (green dashed). The upper panel corresponds to the blue line ($k=2.4$, $\alpha=4.5$) in (a) with probe mass $M=30$, and the lower panel corresponds to the blue line ($k=3.3$, $\alpha=1.6$) in (b) with probe mass $M=40$.
   (d) Diffusion coefficient of the probe for different mass: from the effective active motion (blue line) and the simulation (orange points). The upper panel and lower panel share the same parameters with the blue lines in (a) and (b), respectively.}
   \label{fig:1d}
\end{figure*}

%Although a quantitative description of the dynamics requires detailed knowledge of all terms, the qualitative behavior of the reduced dynamics is closely related to the sign of friction. 

The dynamics of the probe can be classified into different regimes according to the behavior of friction $f(v)$ (or more precisely $f(v)+G(v)/M$); see Figs.~\ref{fig:1d}(a)-(b):  \\ 
(R1) A standard passive regime: $f(v)>0$ for any $v>0$. \\
(R2a) A peculiar active regime: $f(v)<0$ for small $0<v<v^*$ and $f(v)>0$ for large $v>v^*$.
\\
(R2b) Another peculiar active regime: $f(v)>0$ for $0<v<v^\dagger$, $f(v)<0$ for $v^\dagger<v<v^*$, and $f(v)>0$ for $v>v^*$. 

%\st{In the following, we discuss how the probe behaves in the different regimes.} 
In the standard (passive) regime (R1), the probe velocity fluctuates around $0$. 
Expanding $f(v)\sim f^\prime(0) v=\gamma v$ and $B(v)\sim B(0)$,
the dynamics can be described by \eqref{sd}, which is an underdamped passive Brownian motion.

In regime (R2a), 
the stationary velocity distribution is bimodal with peaks around $\pm v^*$, as shown in Fig.~\ref{fig:1d}(c).
Most of the time, the probe moves at velocity around $\pm v^*$.
Occasionally, at random times, the velocity transits between $\pm v^*$.
The transition rate $\alpha^*$ can be obtained from the exact mean first-passage time for \eqref{1dr},
\begin{equation}\label{rate}
    \alpha^{*-1} = \int_{-v^*}^{v^*}\ud y\int_{-\infty}^y\ud z \frac{\exp[\psi(y)-\psi(z)]}{B(z)},
\end{equation}
with effective potential $\psi(v) = \int^v\ud w \left[\frac{Mf(w)+G(w)}{B(w)}\right] $; see SM \cite{SM}.
We may approximate the above formula by the Kramers formula \cite{Moreno2020} if the transitions are rare.
Furthermore, the probe motion \eqref{1dr} in regime (R2a) can be reduced to an underdamped run-and-tumble motion: expanding around $v^*$ and taking into account the transition, we obtain
\begin{equation}\label{1dsim}
    M\dot v= -\mu^*(v-\sigma^* v^* )  +\sqrt{2B^*}\,\xi,\quad \sigma^*= \pm1,
\end{equation}
where $\mu^*=f^\prime(v^*)$ is the effective friction coefficient, 
$\sigma^*v^*$ represents the propulsion velocity of the probe, flipping randomly at rate $\alpha^*$, 
and $\sqrt{2B^*}\xi=\sqrt{2B(v^*)}\xi$ is the translational noise. 
From \eqref{1dsim}, we analytically deduce the diffusion coefficient $D$ of the probe, 
\begin{equation}
    D=\lim_{t\rightarrow\infty}\ev{q(t)^2}/(2t) =v^{*2}/(2\alpha^*) + B^*/\mu^{*2} .
\end{equation}
According to \eqref{rate}, $\alpha^*$ decreases exponentially with the probe mass $M$, and hence, the diffusion is exponentially enhanced for large $M$. 

\replaced{I}{Finally, i}n regime (R2b), 
there are three peaks in the stationary velocity distribution: $0,\pm v^*$.
Two transition rates $\alpha_{0}^*$ (for $\pm v^*\rightarrow0$) and $\alpha_{ 1}^*$ (for $0 \rightarrow\pm v^*$) can also be solved analytically \cite{SM}. 
The dynamics in (R2b) can be further reduced to an underdamped ``run-and-stop motion",
characterized by
\begin{equation}\label{r2b}
    M\dot v= -\mu^*_{\sigma^*}(v-\sigma^* v^* )  +\sqrt{2B^*_{\sigma^*}}\,\xi,\quad \sigma^*= 0,\pm1.
\end{equation}
$\mu^*_{\sigma^*}=f^\prime(\sigma^*v^*)$ and $B^*_{\sigma^*}=B(\sigma^*v^*)$ now depend on the current value of $\sigma^*$.
From this effective active dynamics, the diffusion coefficient is given by 
\begin{equation}
    D=\frac{2 \alpha_{ 1}^*}{\alpha_{0}^* +  2 \alpha_{ 1}^*} \left(\frac{v^{*2}}{\alpha_{0}^*}+\frac{B^*_1}{\mu^{*2}_1}\right)+ \frac{\alpha_{0}^* }{\alpha_{0}^* + 
 2 \alpha_{1}^*}\frac{B^*_0}{\mu_0^2} .
\end{equation}

So far, there are two levels of reduced descriptions,
the reduced nonlinear dynamics \eqref{1dr} and the effective active dynamics \eqref{1dsim}\eqref{r2b}. 
The reduced nonlinear dynamics \eqref{1dr} is completely quantitative and universally valid for all regimes. 
Compared with direct simulations for the composite probe-medium system, we find excellent agreement for both the stationary distributions of the probe velocity $\rho^{\text{st}}(v)$ and the transition rate, respectively shown in Fig.~\ref{fig:1d}(c) and in SM \cite{SM}. 
The effective active dynamics \eqref{1dsim}\eqref{r2b} serves as semi-quantitative approximations in the respective peculiar regimes. 
In velocity space, its prediction of the stationary distribution $\rho^{\text{st}}(v)$ deviates somewhat from the simulations, as shown in Fig.~\ref{fig:1d}(c).
In position space, it quantitatively predicts the diffusion coefficient, in \deleted{excellent }agreement with simulations; see Fig.~\ref{fig:1d}(d). 
This description is useful for an intuitive understanding of the active behavior of the probe \deleted{motion }in regimes (R2a)(R2b).

\begin{figure*}[!htp]
    \centering
    \includegraphics[scale=0.071]{pic2d.pdf}
    \caption{For 2D active Brownian medium with Lennard-Jones interaction: 
    (a) Landscapes of $f(v)$, $B_\parallel(v)$, and $B_\perp(v)$ (per medium particle) for different $\alpha$, showing regimes (A1) and (A2a). Other parameters are $k=2.4$, $u=3$, $R=0.5$, $L=10$, $\mu=1$. 
    (b) Same for $k=1.95$, $\alpha=3.0$, corresponding to regime (A2b). 
    (c) Stationary distributions of the probe velocity $\rho^\text{st}(v_x,v_y)$ and speed $\rho^\text{speed}(v)=2\pi v\rho^\text{st}(v_x,v_y)$. The 3D plots of $\rho^\text{st}(v_x,v_y)$ and the red-dashed lines in $\rho^\text{speed}(v)$ are obtained from the reduced dynamics \eqref{2dr}. The blue line denotes the simulation results of the composite system. The green-dashed line is from the effective active motion. The upper panel corresponds to  $k=2.4$, $\alpha=4.5$ in (a) and $M=30$; the lower corresponds to $k=1.95$, $\alpha =3.0$ in (b) and $M=46$.
    (d) Diffusion coefficient of the probe as a function of mass: from the effective active motion (blue line) and the simulation (orange points). The upper panel and lower panel correspond to $k=2.4$, $\alpha=4.5$ in (a) and $k=1.95$, $\alpha =3.0$ in (b), respectively.}
    \label{fig:2d}
\end{figure*}

\textit{2D active Brownian medium.}
We continue with active Brownian particles \cite{Basu2018} for the medium.
\replaced{For probe velocity}{Supposing the probe velocity is} $(v_x,v_y)$, the dynamics \eqref{medium_dynamics} for one active Brownian particle in the moving frame is then
\begin{equation}\label{2dm}
\begin{aligned}
    \mu\dot r_x &= F(r)\hat r_x + \mu u \cos\phi -\mu v_x,\\
    \mu\dot r_y &= F(r)\hat r_y + \mu u \sin\phi -\mu v_y,\\
    \dot \phi &= \sqrt{2\alpha} \xi,
    \end{aligned}
\end{equation}
with propulsion speed $u$, and $\alpha>0$ characterizing the persistence of the propulsion angle $\phi$.

Adopting polar coordinates in velocity space $(v,\theta)$ \deleted{for the probe, }with $v_x=v\cos\theta, v_y=v\sin\theta$, we obtain the reduced nonlinear dynamics \cite{SM}\replaced{for the probe:}{, given by }
\begin{equation}\label{2dr}
\begin{aligned}
    &M \dot v(t) = - f(v(t)) +\sqrt{2B_\parallel(v(t))} \xi_\parallel(t) \\
    &\quad\quad\quad-\frac 1M G(v(t))+\frac{1-\eta} MB^\prime_\parallel(v(t))+\frac{B_\perp(v(t))}{Mv(t)},\\
    &M v(t) \dot \theta(t) = \sqrt{2B_\perp(v(t))} \xi_\perp(t) ,
    \end{aligned}
\end{equation}
where $\xi_\parallel$ and $\xi_\perp$ are independent standard white noises, respectively causing the fluctuation along and perpendicular to the current velocity. 
$\eta$ depends on the discretization convention, same as in 1D.
The last term in the first equation, $B_\perp/(Mv)$, originates from the use of polar velocity coordinates. 
The nonlinear friction $f(v)$, its second-order correction $G(v)$, and the velocity-dependent noise intensities $B_\perp(v),B_\parallel(v)$ are given explicitly as expectation values of a ``fixed-$\mathbf v$" dynamics; see the End Matter.
Except for $f(v)$ of order $O(\epsilon)$, the other terms are of order $O(\epsilon^2)$.
%\added{The total friction up to $O(\epsilon^2)$ is $g(v)=f(v)+G(v)/M$.}

The nonlinear friction $f(v)$ may take negative values. 
\replaced{T}{However, t}he mechanism in 2D is \replaced{more}{much} complicated than that in 1D since the medium particles can bypass the probe\added{ (see the End Matter)}. \deleted{In the End Matter, we include an intuitive explanation.}
For $f(v)$ to exhibit negative values, firstly, as in 1D,
the persistence length should be large compared to the interaction length. 
Secondly (and different from 1D), the interaction can have a hardcore, but in this case, the force should decay slowly at large distances, instead of being a direct collision.

As an example, we investigate the Lennard-Jones potential, a realistic interaction with a hardcore.
We numerically calculate all quantities and plot $f(v)$, $B_\parallel(v)$, and $B_\perp(v)$ in Figs.~\ref{fig:2d}(a)-(b). 
The Lennard-Jones force is $F(r)=k/k_0[(R/r)^{13}-(R/r)^7]$  with $R$ denoting the size of the probe and $k$ representing the strength of the interaction \footnote{The Lennard-Jones force here uses a different set of parameters compared to its common form. $k_0$ is chosen to be $\frac{42}{169}\sqrt[6]{\frac{7}{13}}$. This choice benefits the comparison with 1D case. Here, $R$ corresponds to zero-force distance, characterizing the size of the probe. $k$ equals the maximum amplitude of the attraction and determines whether the medium particles are trapped by the probe}. $k_0$ is chosen to render $\min_r F(r)=-k$.

Similarly to 1D, there are three distinct regimes determined by the sign of $f(v)$: \\
(A1) A standard passive regime: $f(v)>0$ for all $v>0$. \\
(A2a) A peculiar active regime: $f(v)<0$ for $v\in(0,v^*)$ and $f(v)>0$ for $v>v^*$.\\
(A2b) Another active regime: $f(v)>0$ for $v\in(0,v^\dagger)$, $f(v)<0$ for $v\in(v^\dagger,v^*)$, and $f(v)>0$ for $v>v^*$. 
\deleted{In the active regimes, the probe dynamics is distinct from 1D, which we specify next. }

In the standard (passive) regime (A1), the probe is well described as an underdamped Brownian motion \eqref{sd} with linear friction coefficient $\gamma=f^\prime(0)$ and constant noise intensity $B=B_\parallel(0)=B_\perp(0)$.

In the active regime (A2a),
the probe moves at a speed around $v^*$; see Fig.~\ref{fig:2d}(c). 
Note the significant difference between 1D and 2D. 
The direction angle $\theta$ of the velocity changes continuously and follows a free diffusion.
The dynamics can be further simplified to an active motion.
Expanding $f(v)$ and $B(v)$ around $v^*$, \deleted{we find }the probe follows an effective active motion, described by
\begin{equation}\label{2ds}
    \begin{aligned}
        M\dot v &= -\mu^* (v-v^*)+\frac{B_\perp^*}{Mv}+\sqrt{2B_\parallel^*}\xi_\parallel, \\
        Mv^*\dot \theta &= \sqrt{2B_\perp^*} \xi_\perp.
    \end{aligned}
\end{equation}
This is a special kind of underdamped active Brownian motion, in which
the propulsion direction aligns with the direction of the velocity, sharing a common rotational diffusion process.
$\mu^*=f^\prime(v^*)$ plays the role of friction coefficient, 
$v^*$ represents the propulsion speed, and $B_\perp^*=B_\perp(v^*)$, $B_\parallel^*=B_\parallel(v^*)$ are noise intensities. 
% The stationary distribution of the speed can be obtained analytically as 
% \begin{equation}
%     \rho_v \propto v^\frac{B_\perp(v^*)}{B_\parallel(v^*)}\exp\left[-\frac{Mf^\prime(v^*)}{2B\parallel^*}  (v-v^*)^2\right].
% \end{equation}
From \eqref{2ds}, the diffusion coefficient is obtained as
\begin{equation}
    D=M^2v^{*4}/(2B_\perp^*) +(B_\perp^*+B_\parallel^*)/(2\mu^{*2}).
\end{equation}
Different from 1D, the diffusion is enhanced quadratically as $M$ increase, instead of exponentially.

In the active regime (A2b), \replaced{the}{
ignoring the noise, there are two stable speeds, $0$ and $v^*$. 
The} probe velocity stays around either $v=0$ or $v=v^*$, with random transitions between them; see Fig.~\ref{fig:2d}(c). 
Thus, the probe motion randomly switches between active and passive Brownian motions, described by \eqref{sd} and \eqref{2ds}, respectively.
Two switching rates can also be obtained. 

Again, the nonlinear dynamics \eqref{2dr} is completely quantitative, while the effective active motion \eqref{2ds} is semi-quantitative (qualitative for velocity distribution but quantitative for diffusion coefficient). 
See Fig.~\ref{fig:2d}(c) and Fig.~\ref{fig:2d}(d) for the comparison of velocity distribution and of diffusion coefficient, respectively.

\textit{Experimental proposal.}
Our theory requires an underdamped probe. 
The proper experimental systems can be 
active granular particles (for example,  vibrobots \deleted{in Ref.~}\cite{Caprini2024}) but not bacterial baths.
Moreover, in 2D, a hardcore interaction should decay slowly at large distance to transfer activity, instead of being collisions.
\replaced{One scenario uses a passive probe decorated with silicon gel to soften contact with active granular particles.}{One scenario is to put a passive probe in a medium of active granular particles and to decorate the probe with silicon gel to soften the contact force.}
\replaced{Alternatively, a magnetic probe and ferrous medium particles can provide long-range attraction.}{Meanwhile, we can use a magnetic probe and ferrous medium particles to realize a long-range attractive force.}
The two active regimes (A2a) and (A2b) can be achieved by tuning the relative magnitude between interaction and propulsion force of medium particles.
\deleted{By observing the velocity distribution, we can test whether the transfer of activity scenario works. }

\textit{Conclusion.}
We present a complete solution to \deleted{characterizing }the reduced dynamics for a spherical probe immersed in a scalar active medium and find that
active motion may be transmitted through the induced friction and noise for certain \replaced{couplings}{types of coupling}.  
A related phenomenon was recently reported in a granular matter experiment \cite{Caprini2024}.

% \deleted{Negative nonlinear friction was reported for a penetrable probe in Ref.~\cite{Kim2024}. 
% Going beyond, we have added the fluctuating dynamics by Eq. \eqref{1dr}\eqref{2dr}, and we have identified the active behavior, described by Eqs. \eqref{1dsim}\eqref{r2b}\eqref{2ds}. 
% Furthermore, we find in 2D that the probe does not need to be penetrable, allowing an experimental proposal for our predicted transfer of activity. }

\deleted{Our study may prove constructive for designing artificial micro-devices working in an active (for instance, biological) environment.}
On a more fundamental level, the results are opening
a new avenue for understanding the universal presence of active motion at different scales in nature. To conclude, these findings constitute important evidence for understanding the origin and transfer of active motion, where activity is begotten, not made.  

\begin{acknowledgments}
\textit{Acknowledgments.}
\added{We thank H. T. Quan for helpful comments on the manuscript.} This work received support from the
 China Scholarship Council, No. 202306010398.
\end{acknowledgments}

\bibliography{transmit}

\section{end matter}
\subsection{Expressions of velocity-dependent friction and noise}
In this section, we show the concrete expressions for $f(v)$, $B(v)$, and $G(v)$, respectively in 1D and 2D. 
%We also briefly discuss the relation between them and the fixed-$v$ dynamics. 
%\deleted{They are all expressed as expectation values in the fixed-$v$/fixed-$\mathbf {v}$ dynamics of a single medium particle, which we specify later. }

In 1D, quantities in Eq.~\eqref{1dr} are given by 
\begin{equation}\label{friction}
\begin{aligned}
    &f(v) = N \ev{F(r)}_{v} ,\\
    &B( v)=    N\int_0^\infty \ud s \ev{F(r(s)); F(r(0))}_{ v},\\
    &G( v) = N\int_0^\infty \ud s \ev{F(r(s)); F(r(0))\pdv{v} \log\rho_{ v}(r(0))}_{v}.
    \end{aligned}
\end{equation}
For fixed boundary length $L$, all three quantities are proportional to the number of medium particles $N$. 
Here, we introduce a fixed-$v$ dynamics for a single variable $r$ (relative position of medium particle), given by Eq. \eqref{1dm} with probe velocity $v$ fixed.
In the above formulas,  $\rho_v(r)$ is the stationary distribution in the fixed-$v$ dynamics, $\ev{\dots}_v$ denotes the stationary average, and $\ev{\star;\circ}_v=\ev{\star\circ}_v-\ev{\star}_v\ev{\circ}_v$ denotes the connected correlation (covariance).

In 2D, quantities in Eq.~\eqref{2dr} are given by 
\begin{equation}\label{exp}
    \begin{aligned}
    &f(v) = N\ev{F_\parallel}_{\mathbf v}\\
    &B_\parallel(v) = N\int_0^\infty \ud s \ev{F_\parallel(r(s));F_\parallel(r(0))}_{\mathbf v}, \\ 
    &B_\perp(v) =N \int_0^\infty \ud s \ev{F_\perp(s);F_\perp(0)}_{\mathbf v} ,\\
    &G(v)= N\int_0^\infty\ud s \ev{F_\parallel(r(s));F_\parallel(r(0))\partial_v\log\rho_\mathbf v(r(0))}_\mathbf v.
\end{aligned}
\end{equation}
$F_\parallel$ and $F_\perp$ represent the interaction force in the tangential and perpendicular directions with respect to the probe velocity, respectively.
Similar to 1D, 
the fixed-$\mathbf v$ dynamics for the relative position $\mathbf r$ of a single medium particle is given by Eq.~\eqref{2dm} with fixed probe velocity $(v_x,v_y)$. 
In the above formula, $\ev{\dots}_{\mathbf v}$ and $\rho_\mathbf{v}$ respectively denote the average and the stationary distribution in the fixed-$\mathbf v$ dynamics.  

In both 1D and 2D, 
all quantities can be obtained from the analysis of the fixed-$v$/fixed-$\mathbf {v}$ dynamics, 
which is essentially a nonequilibrium dynamics for a single medium particle. 
They can be calculated or analyzed easily from either theoretic or numerical methods. 

\subsection{Intuitive analysis of nonlinear friction}
Here, we provide a theoretical analysis of $f(v)$ in 1D and 2D.

We specify the setup of our analysis in 1D. 
We consider that a repulsive interaction with a maximum force $F_{\max}=\max_rF(r)$ and an interaction range $R$. 
Hardcore interaction refers to the case of $F_{\max}=\infty$, while soft interaction means that $F_{\max}$ is finite. 
The expression of $f(v)$ in \eqref{friction} involves an average in the fixed-$v$ dynamics. 
In the following, we assume the probe is moving at fixed velocity $v\geq0$ (from left to right) to analyze that average, noting that $f(v)$ is antisymmetric, $f(-v)=-f(v)$.

\begin{figure*}
    \centering
    \includegraphics[page=1, width=0.19\linewidth]{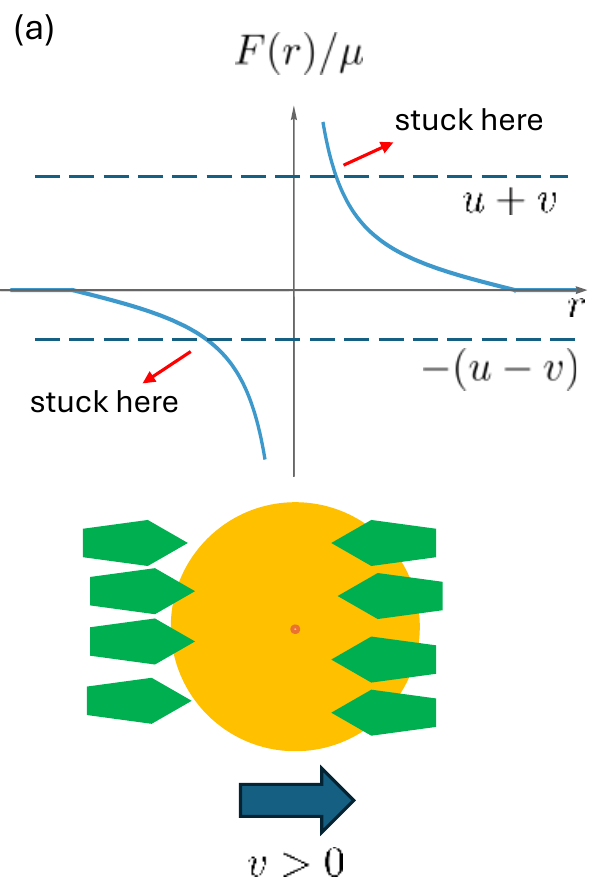}
    \includegraphics[page=2, width=0.19\linewidth]{pic_end.pdf}
    \includegraphics[page=3, width=0.19\linewidth]{pic_end.pdf}
    \includegraphics[page=4, width=0.19\linewidth]{pic_end.pdf}
    \includegraphics[page=5, width=0.19\linewidth]{pic_end.pdf}
    \caption{Mechanism in 1D: The green pentagon represents the medium particles, the acute angles of which denote their propulsion direction. The yellow disk represents the probe, moving rightwards. The blue solid line represents the force (divided by $\mu$) exerted on the medium particles. If medium particles can pass through the probe, the distance (shown in red) between $F(r)/\mu$ and $u+v$ or $-(u-v)$ yields the relative velocity $\abs{\dot r}$ at different position $r$.}
    \label{fig:enter-label}
\end{figure*}

We first consider the large persistence limit of the medium $\alpha R/u\rightarrow0$, where we can assume that half of medium particles have fixed $\sigma=+1$ while the others have fixed $\sigma=-1$.

(0) For hardcore interactions: friction is purely positive. 
%medium particles cannot pass through the probe. 
Medium particles with certain $\sigma$ get stuck at the same distance, as is shown in Fig.~\ref{fig:enter-label}(a). 
Each medium particle with $\sigma = +1$ pushes the probe with force $\mu (u - v)$, 
while particle with $\sigma = -1$ exerts a force $-\mu (u+ v)$. 
The total force \eqref{friction} on the probe becomes $-f(v)=-N\mu v$, indicating a positive and linear friction.

For soft interaction, we further distinguish several cases according to the magnitude of $F_{\max}$. 

(1) Small interaction amplitude $ F_{\max} < \mu u$: the sign structure of $f(v)$ is $(-,+)$, corresponding to the peculiar regime (R2a). 

For small probe velocity $0\leq v<u-F_{\max}/\mu$, the medium particles are able to pass through the probe. 
We consider medium particles with $\sigma=+1$ (moving in the same direction as the probe) 
and $\sigma=-1$ (moving in the opposite direction) separately, noticing that the medium stationary distribution is inversely proportional to the relative velocity $\abs{\dot r}$.
For medium particles with $\sigma=+1$, $\abs{\dot r} = u-v+F(r)/\mu$ is smaller behind the probe ($r<0$), resulting in higher density there (Fig. \ref{fig:enter-label}b).
For medium particles with $\sigma=-1$, $\abs{\dot r} = u+v-F(r)/\mu$ is smaller in front of the probe ($r>0$), resulting in higher density there (Fig. \ref{fig:enter-label}c).
The density asymmetry for $\sigma=+1$ dominates, causing a higher medium density behind the probe than in front of the probe. 
This density difference pushes the probe in the direction of its motion, resulting in negative friction.

% For medium particles with $\sigma=+1$, the relative velocity is given by $\abs{\dot r} = u-v+F(r)/\mu$, which is smaller when medium particles are behind the probe ($r<0$) and is larger when they are in front of the probe ($r>0$). 
% Since the medium stationary distribution is inversely proportional to the relative velocity $\rho_v(r,\sigma=+1)\propto 1/\abs{\dot r}$, the medium density behind the probe is higher than that in front of the probe, as shown in Fig.~\ref{fig:enter-label}(b). 
% On the other hand, for medium particles with $\sigma=-1$, the relative velocity is given by $\abs{\dot r}=u+v-F(r)/\mu$, and the profile of $\rho_v(r,\sigma=-1)$ is reversed: the density behind the probe is lower; see Fig.~\ref{fig:enter-label}(c). 
% The density asymmetry for $\sigma=+1$ dominates, 
% and the total density $\rho_v(r)=\sum_{\sigma=\pm1}\rho(r,\sigma)$ is larger when medium particles are behind the probe. 
% This density difference push the probe in the direction of its motion, resulting in negative friction. 

The above analysis is for small probe velocity. 
For large enough $v$, medium particles with $\sigma=+1$ cannot catch up with the probe, leading to $f(v)>0$ for large $v$.

(2) $F_{\max}\gtrsim\mu u$ (maximum repulsion is a bit larger than propulsion force): the sign structure of $f(v)$ is $(+,-,+)$, corresponding to the peculiar regime (R2b). 

The medium particles get stuck for small $v$, as shown in Fig.~\ref{fig:enter-label}(d). 
Similar to the hardcore case, 
%particles with $\sigma=+1$ have a negative contribution while particles with $\sigma=-1$ have a positive contribution to the friction.
the overall $f(v)$ is positive, linearly increasing with respect to $v$.

Nevertheless, for larger $v$ such that $v+u>F_{\max}/\mu$ (but still $v<u$), medium particles with $\sigma=-1$ are able to pass through the probe while particles with $\sigma=+1$ are still stuck; see Fig.~\ref{fig:enter-label}(e). 
Medium particles with $\sigma=-1$ still have a positive contribution to $f(v)$ but much less than that when they are stuck. 
Therefore, $f(v)$ can decrease to a negative value. 
At the threshold velocity $v+u=F_{\max}/\mu$, there is a singularity in the stationary distribution $\rho_v$, and $f(v)$ drops suddenly, which accounts for the singular turn over in Fig.~\ref{fig:1d}(b).

For very large $v$ such that $v>u$, $f(v)$ becomes positive again.

(3) $F_{\max}\gg\mu u$: the interaction is very large and effectively hardcore. $f(v)$ is purely positive.

For large flip rate $\alpha$ (small persistence), run-and-tumble motion resembles (passive) Brownian motion with diffusivity $D=u^2/(2\alpha)$ \cite{Solon2015}, and the medium reduces to an equilibrium medium. 
For an equilibrium medium, we know that the friction $f(v)$ is always positive.

In summary, for either (effectively) hardcore interaction or large $\alpha$, the behavior of $f(v)$ belongs to the passive regime (R1). 
For soft interaction and small $\alpha $, if $F_{\max}<\mu u$, $f(v)$ behaves as in the active regime (R2a), and if $F_{\max}\gtrsim \mu u$, we get the active regime (R2b). \\

For 2D, the mechanism is much more complicated than that in 1D. 
Medium particles can pass by a hardcore probe. 
We discuss how the negative friction can appear for an interaction with a hardcore. 

We focus on the case of high-persistence limit, $\alpha R/u\rightarrow0$. 
In this intuitive analysis, we roughly divide a hardcore interaction into two parts: 
a core region and a peripheral region.
In the core region,  the repulsive force between the probe and medium particles is large, so that medium particles hardly enter this region.
% When medium particles are close to the probe, the repulsive force is large. Around the probe, there is a circular region that medium particles hardly enter. 
% We call it the core region. 
The peripheral region is outside the core region, where the interaction force is smaller than but comparable to the propulsion force, either repulsive or attractive. 
%Outside the core region, there is a region where the interaction force is smaller than but comparable to the propulsion force, either repulsive or attractive. 
Medium particles can enter and depart this region easily. 
%We call it the peripheral region. 

When a bath particle interacts with the probe, there are two cases. 
In the first case, the bath particle hits the core region and feels a strong repulsion. 
It then slows down and moves along the boundary of core region before leaving. 
This kind of events is similar to the 1D case where bath particles are stuck,
resulting in a positive contribution to the friction. 
In the second case, the bath particle enters the peripheral region but does not hit the core region. 
Since the interaction is not large, the trajectory of the particle only deviates slightly from a straight line during the interaction. 
This kind of events is similar to the 1D case where bath particles pass through the probe. 
It results in a negative friction, pushing the probe in the same direction as its motion. 

The total friction is a competition of the two contributions. 
In order for negative friction, most bath particles should pass through the peripheral region rather than collide with the core region. 
Therefore, the interaction should have a large peripheral region and a small core region.
This means that the interaction decays slowly as the distance increases instead of being direct collision.
In addition, an interaction with attraction at large distance (such as Lennard-Jones) is preferable than a purely repulsive one. 
The attractive part captures a lot of bath particles in the peripheral region, enhancing the negative contribution.
%Unlike the 1D case, we do not have a good intuitive explanation accounting for the regimes (R2a) and (R2b).

In SM \cite{SM}, we provide a set of plots, showing how different interaction strengths of the Lennard-Jones force, by altering the two contributions, lead to different regimes of probe motion in 2D.

\end{document}

\begin{CJK*}{UTF8}{gbsn} % Use default fonts from CJK (see below)

\title{Supplemental Material:\\
Transfer of active motion from medium to probe\\ via the induced friction and noise}

\author{Ji-Hui Pei (裴继辉) \orcidlink{0000-0002-3466-4791}}
\affiliation{School of Physics, Peking University, Beijing, 100871, China}
\author{Christian Maes \orcidlink{0000-0002-0188-697X}}
\affiliation{Department of Physics and Astronomy, KU Leuven, 3000, Belgium}
\email{christian.maes@kuleuven.be}

\maketitle
\end{CJK*}

\section{Derivation of the reduced dynamics}
In this section, we show how to integrate out the medium particles, resulting in a generally valid reduced dynamics that contains velocity-dependent friction and noise.

We recall the setup where an inertial spherical particle, called probe, is coupled to a homogeneous, dilute medium composed of $N$ independent active particles.  
One important theoretical idea is to adopt a time-dependent reference frame where the probe remains at the origin. 
By doing so, we eliminate the probe position, and we avoid the problem of its possible fast change that would undermine the time-scale-separation assumption.

The evolution equation is given by Eq. (3) in the main text, which is repeated here, 
\begin{align}
        \mu\dot r^a_i &=-\mu v_i +F_i(\vec r^a) +A_i^a ,\label{medium_dynamics}\\
        M\dot v_i &= \sum_{a=1}^N -F_i(\vec r^a) ,
\end{align}
where the subscript $i$ denotes the spatial components of vectors. 
In the above equations, the probe velocity $\vec v$ changes slowly compared to the motion of $\vec r^a$. 
Let $\epsilon>0$ be small, to characterize the dimensionless ratio of time-scales between $\vec r^a$ and $\vec v$. 
The analysis of the time scales and the determination of $\epsilon$ are given in Sec. \ref{scale}. 

It is now possible to integrate out the fast variables $\vec r^a$ by taking advantage of time-scale separation. 
In particular, we adopt the projection-operator method, a systematic way of quasistatic expansion, described in Refs. \cite{Solon2022,Pei2025}. Please see Eqs. (1)-(4) in Ref. \cite{Pei2025} for the general formulism.

By applying that quasistatic expansion, one introduces a ``fixed-$\vec v$" dynamics for a single medium particle (only), where $\vec {r}^a_i$ evolve according to Eq.~\eqref{medium_dynamics} with fixed $\vec v$. 
The stationary distribution of the fixed-$\vec v$ dynamics is denoted by $\rho_{\vec v}(\vec r^a)$. The corresponding average and covariance are respectively denoted by $\ev{\cdots}_{\vec v}$ and  $\ev{\circ;\star}_{\vec v}=\ev{\circ\,\star}_{\vec v}-\ev{\circ}_{\vec v}\ev{\star}_{\vec v}$.

Up to the second order in $\epsilon$, the reduced dynamics for the probe is described by a 
Fokker-Planck equation for the velocity distribution function (of the probe) $\tilde\rho(\vec v,t)$,
\begin{equation*}
    M^2\pdv{t}\tilde\rho(\vec v,t)= \sum_i \pdv{v_i}\left[(M f_i(\vec v)+G_i(\vec v))\tilde\rho(\vec v,t)\right]-\sum_i \pdv{v_i}\left[\sum_j \pdv{B_{ij}(\vec v)}{v_j}\tilde\rho(\vec v,t)\right]+\sum_{ij}\pdv{v_i}\pdv{v_j} \left[ B_{ij}(\vec v)\tilde\rho(\vec v,t)\right].
\end{equation*}
The corresponding 
stochastic differential equation is, 
\begin{equation}\label{uniform}
    M\dot v_i(t) = -f_i(\vec{v}(t)) - \frac 1{M}G_i(\vec{v}(t)) +\frac{1}{M}\sum_j \pdv{B_{ij}(\vec{v}(t))}{v_j} -\frac{\eta}{M}\sum_{jk}\pdv{\sigma_{ik}}{v_j}\sigma_{jk}(\vec{v}(t)) +  \sum_j \sigma_{ij}(\vec{v}(t))  \,\xi_j(t) ,
\end{equation}
where the $\xi_j(t)$ are independent standard white noises, and
$\eta$ depends on the discretization convention: $\eta=0$ for the It\^o, $\eta=1/2$ for the Stratononvich, and $\eta=1$ for the anti-It\^o convention. 
On the right-hand side in the above equation, only the first term, $f_i(\vec v)$, is of the order $O(\epsilon)$. Other terms are all of the order $O(\epsilon^2)$. 
The (first-order) nonlinear friction is 
\begin{equation}\label{non_fri}
    f_i(\vec{v}) = N\,\ev{F_i(\vec r)}_{\vec v} .
\end{equation}
The second-order correction of the friction is given by
\begin{equation}\label{additional}
    G_i(\vec v) = N\int_0^\infty \ud s \,\ev{F_i(\vec r(s)); \sum_{j=1}^d F_j(\vec r(0))\pdv{v_j} \log\rho_{\vec v}(\vec r(0))}_{\vec v} .
\end{equation}
The matrix $B_{ij}$ is
\begin{equation}\label{noise_matrix}
    B_{ij}(\vec v)=    N\int_0^\infty \ud s \ev{F_i(\vec r(s)); F_j(\vec r(0))}_{\vec v}.
\end{equation}
with noise coefficients $\sigma_{ij}$ given through
\begin{equation}\label{noise_coefficient}
    \sum_k\sigma_{ik}(\vec v)\sigma_{jk}(\vec v) = B_{ij}(\vec v)+B_{ji}(\vec v) .
\end{equation}
The matrix $\sigma_{ij}$ are not uniquely defined, but different choices lead to the same dynamics as long as the above relation is satisfied. A convenient choice is  $\sigma=\sqrt{B+B^T}$. % (defined by taking the square root of elements in the frame where $B+B^T$ is diagnal). 

In 1D, the dynamics reduce to 
\begin{equation}\label{exp1d}
    M\dot{v}(t) = -f (v(t))  -\frac 1M G(v(t)) + \frac{1-\eta}{M} B^\prime(v(t)) +\sqrt{2B(v(t))}\,\xi(t),
\end{equation}
which is Eq. (5) of the main text. 
%In 2D, we will illustrate how Eq.~\eqref{uniform} is further simplified in the next section. 

% In 1D, we use $v$ to denote the velocity which can be negative (while in other dimensions $v=\abs{\vec v}$ indicates the amplitude of the velocity).

% All quantities in Eq.~\eqref{uniform} reduce to scalars, and 
% the resulting reduced dynamics is 
% \begin{equation}
%     M\dot{v}(t) = -f (v(t))  -\frac 1M G(v(t)) + \frac{1-\eta}{M} B^\prime(v(t)) +\sqrt{2B(v(t))}\,\xi(t).
% \end{equation}
% Here, quantities are given by 
% \begin{equation}
% \begin{split}\label{exp1d}
%     f(v) &= N \ev{F(r)}_{v} ,\\
%     B( v)&=    N\int_0^\infty \ud s \ev{F(r(s)); F(r(0))}_{ v},\\
%     G( v) &= N\int_0^\infty \ud s \ev{F(r(s)); F(r(0))\pdv{v} \log\rho_{ v}(r(0))}_{v} .
% \end{split}
% \end{equation}
% These are Eqs.~(5) and (11) of the main text. 

\section{Dynamics in polar coordinates}
In this section, we illustrate how to rewrite the dynamics in polar coordinates (of the velocity) in 2D, which simplifies the result.

We denote $\vec v = (v_x,v_y)$.
The dynamics in Cartesian coordinate is given by Eqs. \eqref{uniform}-\eqref{noise_coefficient}.
In general, $B_{ij}$, as well as $\sigma_{ij}$, is not a diagonal matrix, meaning that the noises in the $x$-direction and $y$-direction are generally not independent.
Benefiting from the reflection symmetry and the rotation symmetry, we can deduce some properties of $f_i(\vec v)$, $G_i(\vec v)$, and $B_{ij}(\vec v)$. 
Without loss of generality, we suppose the velocity $\vec v=v \vec e_x$ is along the $x$-direction.
Then, the nonlinear friction $\vec f(v\vec e_x)$ is along the $x$-direction, and the perpendicular component vanishes since $\ev{F_y}_{v\vec e_x}=0$ due to the reflection symmetry. 
Also, the second order correction $G_i$ only has a $x$-direction component $G_x$, and its expression is simplified because only the $j=x$ term survives in the summation over $j$ on the right-hand side of Eq.~\eqref{additional}.
Finally, the noise matrix $B_{ij}$ is now diagonal (only if $\vec v$ is along $x$-direction) since $\ev{F_{y}(s);F_{x}}_{v \vec e_x}=\ev{F_{x}(s);F_{y}}_{v \vec e_x}=0$.

We employ polar coordinates in the velocity space $(v,\theta)$, which satisfy $v_x=v\cos\theta$ and $v_y=v\sin\theta$. 
This also brings two orthogonal bases, $\vec e_\parallel=(\cos\theta,\sin\theta)$ along the current velocity and $\vec e_\perp=(-\sin\theta,\cos\theta)$ perpendicular to the velocity. 
A vector $\vec w$ is decomposed as $\vec w = w_\parallel \vec e_\parallel+w_\perp\vec e_\perp$.

%$v(t)$ represents the probe speed at time $t$, and $\theta(t)$
Using the It\^o lemma, we can rewrite the reduced dynamics in the (velocity-space) polar coordinates, %written as 
\begin{equation}\label{decomposition}
\begin{aligned}
    &M \dot v(t) = - f(v(t)) +\sqrt{2B_\parallel(v(t))}\, \xi_\parallel(t) -\frac 1M G(v(t))+\frac{1-\eta} MB^\prime_\parallel(v(t))+\frac{B_\perp(v(t))}{Mv(t)},\\
    &M v(t)\, \dot \theta(t) = \sqrt{2B_\perp(v(t))}\, \xi_\perp(t) ,
    \end{aligned}
\end{equation}
where 
$\xi_\parallel$ and $\xi_\perp$ are now independent standard white noises with intensities $B_\parallel(v)$ and $B_\perp(v)$, respectively. 
The term $B_\perp/(Mv)$ in \eqref{decomposition} comes from the coordinate transformation, which plays an important role when $v$ is very small. 
%Again, only $f(v)$ is of order $O(\epsilon)$. All other terms are of order $O(\epsilon^2)$. 
Expressions are given by
\begin{equation}\label{exp2d}
\begin{aligned}
    &f(v) = N\ev{F_\parallel}_{\vec v},\\
    &B_\parallel(v) = N\int_0^\infty \ud s \ev{F_\parallel(s);F_\parallel(0)}_{\vec v}, \quad B_\perp(v) =N \int_0^\infty \ud s \ev{F_\perp(s);F_\perp(0)}_{\vec v} ,\\
    &G(v) = N\int_0^\infty \ud s \ev{F_\parallel(s);F_\parallel(0)\pdv{v}\log\rho_{\vec v}(0)}_{\vec v} .
\end{aligned}
\end{equation}
Note that although the ensemble average $\ev{\cdots}_{\vec{v}}$ of a general function depends on the direction of $\vec v$, the quantities given in the above formula are only functions of the amplitude $v$, due to their symmetry.

\section{Analysis of the time scales}\label{scale}
The derivation of the reduced dynamics relies on the approximation of time-scale separation. 
However, completely identifying the time scales is not an easy task
because motion constantly changes in the evolution. 
Here, we provide an analysis on how to characterize the time scales, and when the time-scale separation is satisfied. 
%For simplicity, 1D notation is adopted here.

Intuitively, time-scale separation means that the active medium relaxes fast, so that the probe feels that the force comes from a stationary medium. 
%More precisely, we require that the medium relaxation for fixed probe velocity is fast. 
This is related to the time correlation function of interaction forces in fixed-$\vec v$ dynamics,
\begin{equation}
    \ev{\delta\vec F(s)\cdot\delta\vec F(0)}_{\vec v} =K_v(s).
\end{equation}
Its decay rate determines the time scale of the medium particles $\tau_r(v)$, 
% \begin{equation}
%     \tau_r(v)=\dv{s}K_v(s)/K_v(s)\eval_{s=0}, 
% \end{equation}
and this time scale depends on the probe velocity $ v$. 
Within time period $\tau_r(v)$, the correlation decays, and the force on the probe is similar to what comes from a stationary medium with distribution $\rho_{\vec v}$. 

We can analyze $\tau_r(v)$ in the current setup.
The decay of $K(s)$ contains two ingredients. 
One is related to the relaxation of the propulsion direction, characterized by $ 1/\alpha$. 
The other is related to the relaxation of the coordinate, characterized by $R/u$ (interaction length over propulsion speed) which is the typical time for a bath particle to pass through the interaction range. 
Therefore, the time scale is roughly 
\begin{equation}
    \tau_r= \max(\alpha^{-1},R/u).
\end{equation}
In this estimation, $\tau_r$ is independent of $v$.

On the other hand, the time scale of the probe velocity is determined after the reduced dynamics is obtained. 
$\tau_v$ is related to the change rate of $\vec v$, i.e., the acceleration. 
When taking into account only the first-order part of the reduced dynamics, 
we have the equation for the acceleration
\begin{equation}\label{acceleration}
    M \dv{t} \vec v = - f(v) \hat v, 
\end{equation}
with the friction $f(v)=N\ev{F}_v$. 
Meanwhile, we need to figure out how the change of $v$ influences the dynamics of the medium particles. 
We recall the equation of motion for one medium particle, 
\begin{equation}
    \mu\dot{\vec r} = -\mu \vec v +\vec F(r)+\vec A,
\end{equation}
where the right-hand side denotes the total force on the medium particle. 
Its typical value is dominated by the active force $\abs{\vec A}=\mu u$.
Supposing the acceleration of the probe is $\dot{\vec v}$, the change rate of the total force associated with $\dot{\vec v}$ is $\mu \dot{\vec v}$. 
The time scale of the probe is given by the ratio 
\begin{equation}
    \tau_v(v)= \mu u/\abs{\mu\dot{\vec v}}=\frac{Mu}{\abs{f(v)}} =\frac{Mu}{N\abs{\ev{F}_v}},
\end{equation}
which characterizes how fast the change of $v$ influences the medium dynamics. 
In the above formula, we have used the equation of the acceleration \eqref{acceleration}. 
$\tau_v(v)$ depends explicitly on the current velocity of the probe.

Therefore, the ratio of the time scales is 
\begin{equation}
    \epsilon(v) = \frac{\tau_r}{\tau_v} =\max\left( \frac{NR\abs{\ev{F}_v}}{M u^2},\frac{N\abs{\ev{F}_v}}{\alpha M u}\right)
\end{equation}
Note that $\epsilon(v)$ depends on the probe velocity $v$. 
If $\epsilon(v)\ll 1$ is satisfied for a range of $v$, it validates the reduced dynamics for this range of probe velocity. 
If
\begin{equation}
    \epsilon = \max_v \epsilon(v)\ll 1,
\end{equation}
the time-scale separation is valid for arbitrary probe velocity.

There are several ways to validate the time-scale separation: 
First, the mass of the probe $M$ is large. 
Second, the density of the bath is low ($N$ is small). 
Third, the propulsion speed of the medium particles $u$ is large. 
They all lead to a small $\epsilon$.

\section{From nonlinear to linear friction}\label{linn}
%\added{For simplicity, we consider the one-dimensional case only, while }
We show that the nonlinear friction in \eqref{non_fri} verifies $f'(v)\eval_{v=0}= \gamma$, where $\gamma$ is the linear friction coefficient given in Refs.~\cite{Granek2022,Solon2022,Tanogami2022,Pei2025}. 
For simplicity of notation, we consider the 1D case, but the same derivation equally applies to higher dimensions.

In Refs.~\cite{Granek2022,Solon2022,Tanogami2022,Pei2025}, the expression for the linear friction coefficient is 
\begin{equation}\label{fri}
\gamma = -N\int_0^\infty \ud s \ev{F(z(s)-q);\partial_{ q} \log\rho_{ q}^*(z(0))}_{ q}^* .
\end{equation}
There, one introduces another ``fixed-$q$" dynamics (denoted by $*$) for the position of a medium particle $z$, where the frame transformation is not involved, and the probe position is fixed at $ q$ (the probe velocity is therefore zero). 
That is, the dynamics is given by the original composite dynamics (Eq. (2) in main text) with fixed $ q$. 
It should be distinguished with the fixed-$v$ dynamics described previously.
$\rho_{q}^*(x)$ represents the stationary distribution in this fixed-$q$ dynamics, and $\ev{\cdots}^*_q$ denotes the ensemble average. 
We denote the forward generator of the fixed-$q$ dynamics by $L^{*\dagger}_q$.
We can rewrite \eqref{fri} using $L^{*\dagger}_q$,
\begin{equation}\label{gamma/N}
    \begin{aligned}
    \gamma/N    &=-\int_0^\infty \ud s \int\ud z \,\ue^{sL^{*}_q} F(z) \pdv{q} \rho_{q}^*(z)\\
    &=-\int_0^\infty \ud s \int\ud z \,F(z) \ue^{sL^{*\dagger}_q}\pdv{q} \rho_{q}^*(z)\\
    &=-\int\ud z F(z) \frac{1}{L^{*\dagger}_q} \pdv{q} \rho_{q}^*(z) .
\end{aligned}
\end{equation}
Here, the operator (pseudo) inverse $1/L^{*\dagger}_q$ is well-defined,  
$$
    \frac{1}{L^{*\dagger}_q} \nu (x)= -\int_0^\infty \ud s \,\ue^{sL^{*\dagger}_q}\nu(x),$$ 
if  $\nu$ is a traceless function, $\int\ud x\, \nu(x)=0$.

% On the other hand, the derivative of the nonlinear friction $f(v)$ in Eq.~\eqref{non_fri} is 
% \begin{equation}\label{f/N}
%     \begin{aligned}
%     \frac 1{N}\,f'(0) &= \pdv{v} \ev{F(r)}_{v}\eval_{v=0}\\
%     &=\lim_{\delta v\rightarrow0}\int\ud r F(r) [\rho_{\delta v}(r)-\rho_{v=0}(r)]/\delta v
% \end{aligned}
% \end{equation}
%We have used that $q^*-x^*$ is independent of $u$. 
%$\rho_{\delta v}$ here is the stationary distribution of the medium in the moving frame with probe velocity $\delta v$. 
We further construct a time-dependent dynamics (denoted by $\diamond$) for a medium particle: we do not apply the frame transformation but introduce a linearly changing probe position $q^\diamond(t) := q_0+t\delta v$. 
To be specific, in this dynamics, the medium particle position $z$ evolves according to the original composite dynamics (Eq. (2) in the main text) with a predetermined $q^\diamond(t)$. 
The corresponding time-dependent distribution is denoted as $\rho^\diamond(z,t)$. 
We should now distinguish three different distributions: $\rho^*_q(z)$, $\rho_{v}(r)$, and $\rho^\diamond(z,t)$.
In the long-time limit, this time-dependent dynamics distribution reaches a stationary distribution moving at constant velocity $\delta v$, 
and it has the following relation with the fixed-$v$ dynamics $\rho_{\delta v}(r)$ in the moving frame: $\rho^\diamond(r+q^\diamond(t),t)=\rho_{\delta v}(r)$.
Therefore, the nonlinear friction can be expressed as 
\begin{equation}\label{tli}
    \frac1Nf(\delta v)=\int\ud r F(r) \rho_{\delta v}(r)=\int\ud z F(z-q^\diamond(t)) \rho^\diamond(z,t).
\end{equation}
For a small $\delta v$, $\rho^\diamond(z,t)$ evolves mainly along the transient steady distribution $\rho^*_{q^\diamond(t)}(z)$ but with a small correction of the order $O(\delta v)$. Following the quasistatic approximation illustrated in Refs.~\cite{Cavina2017,Khodabandehlou2023}, we have 
\begin{equation}
    \rho^\diamond(z,t)=\rho_{q_0+t\delta v }^*(z)+\delta v \frac{1}{L^{*\dagger}_q} \pdv{q} \rho_{q}^*(z)\eval_{q=q_0+t\delta v} +O(\delta v^2).
\end{equation}
Therefore, applying the above formula to \eqref{tli} yields
\begin{equation}
    \begin{aligned}
    \frac1Nf(\delta v)=\int\ud z F(z-q^\diamond(t)) \rho_{q^\diamond(t)}^*(z) + \delta v \int\ud z \,F(z-q^\diamond(t)) \frac{1}{L^{*\dagger}_{q^\diamond(t)}} \pdv{q^\diamond(t)} \rho_{q^\diamond(t)}^*(z)+O(\delta v^2).
    \end{aligned}
\end{equation}
The above relation holds for any large $t$.
By noticing the translational symmetry of $\rho_q^*(z)=\rho_0^*(z-q)$, we find the first term on the right-hand side vanishes, 
$
    \int\ud x F(z-q^\diamond(t)) \rho_{q^\diamond(t)}^*(z)=f(0)=0.
$
Therefore,
\begin{equation}\label{ff}
    f^\prime(0)  =\lim_{\delta v\rightarrow0} f(\delta v)/\delta v=N \int\ud z \,F(z-q) \frac{1}{L^{*\dagger}_q} \pdv{q} \rho_{q}^*(z) . 
\end{equation}
According to Eqs.~\eqref{gamma/N}\eqref{ff}, we find  
\begin{equation}
    f'(v)\eval_{v=0} =N\int\ud z\, F(z-q) \frac{1}{L^{*\dagger}_q} \pdv{q} \rho_{q}^*(z)=\gamma .
\end{equation}
The derivative of the nonlinear friction $f(v)$ indeed reduces to the linear friction coefficient $\gamma$ appeared in previous studies.

\section{Transition rates}
In this section, we investigate the transition rates that appear in active regimes. 
We illustrate the theoretical predictions for these rates and compare them to the simulation results.

In active regime (R2a), the probe behaves as a 1D underdamped run-and-tumble particle. 
There are two peaks $\pm v^*$ in the velocity distribution $\rho^\text{st}(v)$.
We would like to determine the tumble rate $\alpha^*$ between $\pm v^*$. 
We denote the mean first-passage time  for $v$ to pass from $-v^*$ to $v^*$ as $T_{-v^*\rightarrow v^*}$. 
The tumble rate $\alpha^*$ is given by the inverse of $T_{-v^*\rightarrow v^*}$.
Following the procedure in \cite{Gardiner2009}, we can  exactly solve the mean first-passage time for the nonlinear dynamics \eqref{exp1d}.
The result is 
\begin{equation}
    \alpha^{*-1} = T_{-v^*\rightarrow v^*} = \int_{-v^*}^{v^*}\ud y\int_{-\infty}^y\ud z \frac{\exp[\psi(y)-\psi(z)]}{B(z)},
\end{equation}
with the effective potential 
\begin{equation}
    \psi(v) = \int^v\ud w \left[\frac{Mf(w)+G(w)}{B(w)}\right] .
\end{equation}
In the above expression, $G(v)$ cannot be neglected even in the limit of $M\rightarrow\infty$. 
Therefore, the entire landscapes of $f$, $G$, and $B$ are needed to quantitatively determine $\alpha^*$.
Meanwhile, from the above two equations, we find that the transition rate is depressed exponentially for large mass $M$. 
In the limit of very rare transitions, which is fulfilled if the noise is weak enough, the above tumble rate can be approximated by the Kramers formula:
\begin{equation}\label{Kramers}
    \alpha^* = B(v_\text{max})\frac{\sqrt{\abs{\psi^{\prime\prime}(v_\text{max})\psi^{\prime\prime}(v_\text{min})}}}{2\pi }\exp\left[-\psi(v_{\max})+\psi(v_{\min})\right],
\end{equation}
where $v_\text{max}=0$ and $v_{\min}=v^*$ are the extreme points of the effective potential, and $\psi^{\prime\prime}$ denotes the second-order derivative.

\begin{figure*}[htp]
   \centering
   \includegraphics[scale=0.1]{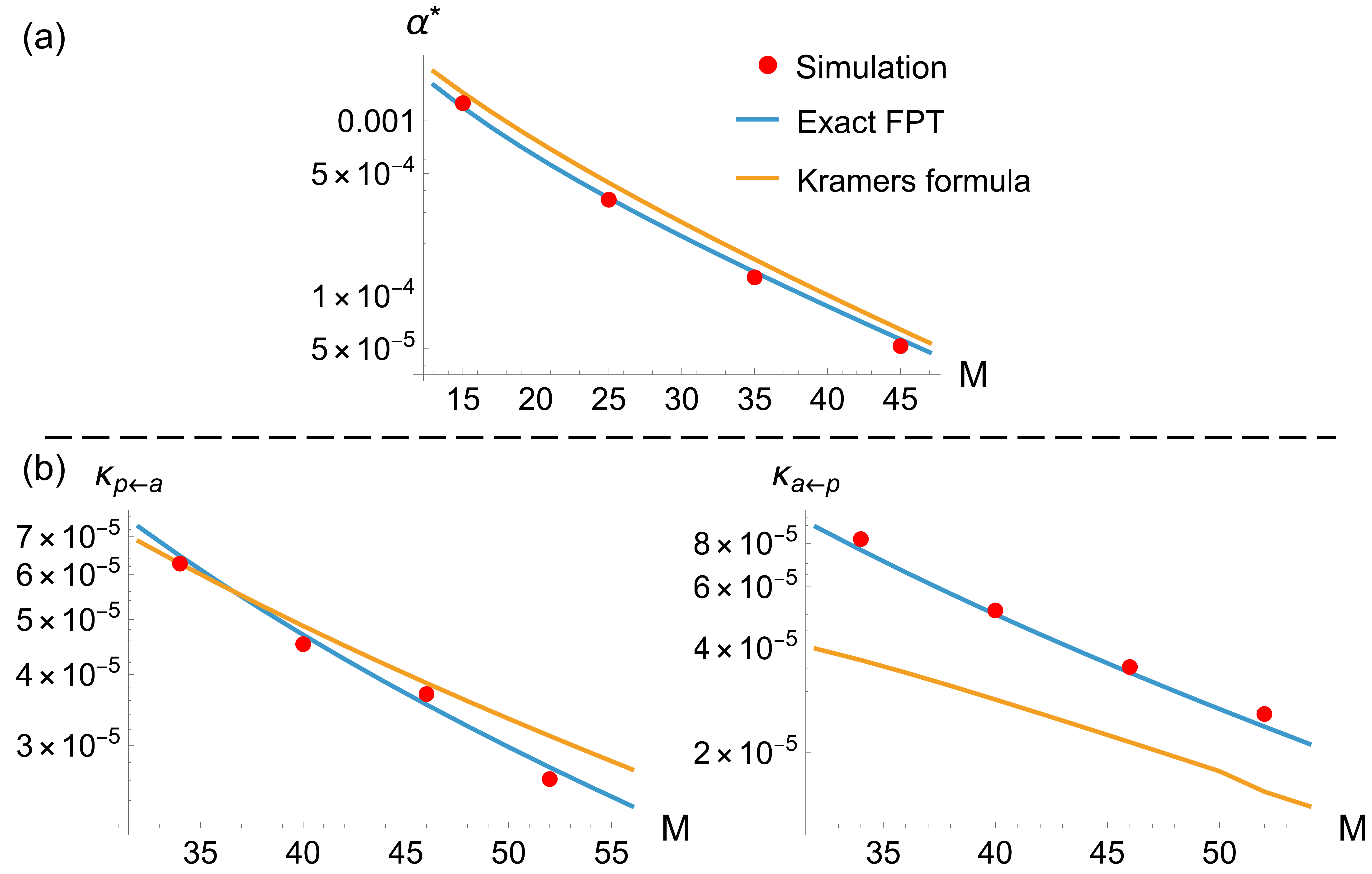}
   \caption{
   (a) In 1D regime (R2a), tumble rate $\alpha^*$ of the probe for different probe masses. The blue and orange lines are from solving first-passage time problem and from the Kramers formula, respectively. The red dots denote the rate observed from simulations. Other parameters are $L=10$, $\alpha =4.5$, $u=3$, $k=2.4$, $\mu=1$, $N=5$, corresponding to the blue line in Fig. 1(a) of the main text. (b) In 2D regime (A2b), two transition rates between passive and active Brownian motion $\kappa_\mathrm{p\leftarrow a}$ and $\kappa_\mathrm{a\leftarrow p}$ for different probe masses. Other parameters are $L=10$, $\alpha=3$, $u=3$, $k=1.95$, $\mu=1$, $N=5$, corresponding to Fig. 2(b) of the main text.. }
   \label{fig:rate}
\end{figure*}

In Fig.~\ref{fig:rate}(a), 
we present a comparison among the tumble rate $\alpha^*$ obtained from first-passage time, that predicted by the Kramers approximation, and that observed in simulations; they agree rather well. 

In active regime (R2b), the probe follows a ``run-and-stop" motion. 
There are three peaks $0,\pm v^*$ in the velocity distribution $\rho^\text{st}(v)$.
The transition rate from ``run" to ``stop" (from $\pm v^*$ to $0$) is denoted by $\alpha_0^*$ while 
the transition rate from ``stop" to ``run" (from $0$ to $\pm v^*$) is denoted by $\alpha_1^*$. 
The velocity cannot directly transit between $\pm v^*$. 
Two different transition rates $\alpha_0^*,\alpha_1^*$ can also be solved from the first-passage time problem.
We should take care of the boundary conditions when solving the first-passage time problem \cite{Gardiner2009}. 
This is because for $\alpha_1^*$, 
there are two directions for $v$ to leave the region around $v=0$, while 
for $\alpha_0^*$,
there is only one direction for $v$ to leave the region around $v=v^*$ or $v=-v^*$. 

The results are given by 
\begin{equation}
    (\alpha^{*}_0)^{-1} = \int_{-v^*}^{0}\ud y\int_{-\infty}^y\ud z \frac{\exp[\psi(y)-\psi(z)]}{B(z)},
\end{equation}
and
\begin{equation}
    (\alpha^{*}_1)^{-1} = \frac12\int_{0}^{v^*}\ud y\int_{-v^*}^y\ud z \frac{\exp[\psi(y)-\psi(z)]}{B(z)}-\frac12\int_{-v^*}^{0}\ud y\int_{-v^*}^y\ud z \frac{\exp[\psi(y)-\psi(z)]}{B(z)},
\end{equation}
where $\psi(v)$ represents the effective potential. 
We also need the full landscape of $f$, $B$, and $G$ to quantitatively determine the rates.
In the limit of rare transition, the above two rates can be approximated by the Kramers formula (Eq. \eqref{Kramers}).

We now come to the active regime (A2b) in 2D, where the probe motion randomly switches between passive Brownian motion and active Brownian motion. 
There are two switching rates: the rate from passive to active Brownian motion $\kappa_\mathrm{a\leftarrow p}$ and the rate from active to passive Brownian motion $\kappa_\mathrm{p\leftarrow a}$. 
They correspond to the transition rates between two peaks in the stationary distribution of the speed. 

We notice that the dynamics of the direction $\theta$ and the dynamics of the speed $v$ are decoupled in Eq. \eqref{decomposition}. 
Therefore, we only need to focus on the dynamics of $v$ and deal with an effective 1D problem. 
The stationary distribution of speed is given by
\begin{equation}
    \rho^\text{speed}(v)\propto\exp\left[-\int^v\ud w\frac{Mf(w)+G(w)-B_\perp(w)/w}{B_\parallel(w)}\right],
\end{equation}
from which we define an effective 1D potential 
\begin{equation}
    \psi_\text{ef}(v) = \int^v\ud w\frac{Mf(w)+G(w)-B_\perp(w)/w}{B_\parallel(w)},
\end{equation}
where we see an extra term related to $B_\perp(w)/w$. 
We denote two local minimum points of $\psi_\text{ef}$ as $v_p$ and $v_a$ ($v_p<v_a$).
Because of the extra term, we find $v_p\neq0$ and $v_a\neq v^*$. 
The transition rates can be obtained from solving the mean first-passage time between $v_p$ and $v_a$. 

The results are given by
\begin{equation}
    \kappa_\mathrm{a\leftarrow p} = \int_{v_p}^{v_a}\ud y\int_{0}^y\ud z \frac{\exp[\psi_\text{ef}(y)-\psi_\text{ef}(z)]}{B(z)},
\end{equation}
\begin{equation}
    \kappa_\mathrm{a\leftarrow p} = \int_{v_a}^{v_p}\ud y\int_{+\infty}^y\ud z \frac{\exp[\psi_\text{ef}(y)-\psi_\text{ef}(z)]}{B(z)}.
\end{equation}
According to the above equations and the expression of $\psi_\text{ef}$, we need the full landscapes of $f$, $B_\parallel$, $B_\perp$, and $G$ to quantitatively determine the rates in 2D.
It is also possible to approximate the above expressions by the Kramers formula.

In Fig.~\ref{fig:rate}(b), 
we compare two rates obtained from the theory with those observed in simulations. 
It turns out that the prediction from the exact first-passage time still matches the simulations very well, 
but the prediction from the Kramers approximation deviates from the simulations. 
This is because the condition of rare transition is not well satisfied: 
see the speed distribution in the lower panel in Fig. 2(c) of the main text, where the height of the valley between two peaks is not low enough. 
In this case, we need to solve the first-passage time to obtain a quantitative prediction instead of using Kramers approximation. 

% \begin{figure}[htp]
%    \centering
%    \includegraphics[scale=0.085]{rate2.pdf}
%    \caption{2D active Brownian medium: 
%    (b) Transition rates between passive and active Brownian motion $\kappa_\text{pa}$ and $\kappa_\text{ap}$ for several different probe masses $M=34,40,46,52$. Other parameters are same as in (a). Other parameters are $L=10$, $\alpha=3$, $u=3$, $k=1.95$, $\mu=1$, $N=5$. }
%    \label{fig:2d}
% \end{figure}

\section{Dependence of Friction on Medium Persistence and Interaction Strength}
This section provides supplementary plots of the nonlinear friction 
$f(v)$ and investigates its dependence on the persistence of the active medium and the strength of the probe-medium interaction.

% If we assume that the velocity of the probe does not change during the interaction. When medium particles pass through the probe, 
% there is a density difference between the medium density behind the probe and that in front of the probe. 
% This density difference pushes the probe in the same direction of the current motion, 
% resulting in negative friction. 

The negativity of friction is a direct cause of the emergent active behavior of probe.
The sign structure of $f(v)$ determines different dynamical regimes of the probe motion reported in the main text.
% The theoretical analysis of negative friction is based on considering a probe moving at a constant velocity $v$ and estimating $f(v)=\ev{F}_v$. 
% Here, we briefly summarize the results and physical picture of the analysis; details are specified in the End Matter. 
Our theoretical analysis, detailed in the End Matter, considers a probe moving at a constant velocity $v$ and estimates the nonlinear friction $f(v)=\ev{F}_v$. 
Below, we summarize the key physical insights from this analysis.

First, in both 1D and 2D, the appearance of negative friction requires that 
the persistence length is large compared to the interaction range ($u/\alpha\gg R$), 
ensuring that the propulsion direction of a medium particle hardly change during an interaction event. 

For such a highly persistent medium, the physical picture is as follows: \\
---In 1D, for a soft repulsive interaction, when medium particles can pass through the probe, the medium density behind the probe can be larger than that in front of the probe, resulting in a negative friction.
Otherwise, if medium particles are stuck by the probe, the friction is positive.
 \\
---In 2D, for interaction with a hardcore and a soft periphery (where interaction is relatively small), total friction
results from the competition between: (1) Positive contributions from medium particles colliding with the hardcore. (2) Negative contributions from medium particles passing through the soft periphery.

% From the about analysis, the persistence and the interaction strength 
% are two key factors that influence the sign of nonlinear friction. 
% In the following, we present several sets of plots of $f(v)$ as the persistence or the interaction strength changes. 

Guided by this picture, we now present numerical results illustrating how 
$f(v)$ evolves with the medium persistence  and the interaction strength. 

We focus on two concrete models described in the main text.
For 1D run-and-tumble medium, the interaction force is a truncated sine function,
$F(r)=k\sin (\pi r/R)$ for $\abs {r}<R$, and the maximum repulsive interaction is $F_\text{max} =k$.
For 2D active Brownian medium, the interaction is given by Lennard-Jones force, 
$F_\text{LJ}(r)=k/k_0 [(R/r)^{13}-(R/r)^{7}]$ with $k_0=\frac{42}{169}\sqrt[6]{\frac{7}{13}}$. 
$k$ denotes the maximum attractive force. 
The Lennard-Jones force has both a hardcore region where the repulsion is very large and a soft region where the interaction is attractive and decays slowly.

\subsection{Dependence on Medium Persistence}
The persistence of the medium active particles is characterized by $\alpha$, which represents the tumbling rate for 1D run-and-tumble particles and the rotational diffusion coefficient for 2D active Brownian particles.
%This persistence reflects how far the medium is from equilibrium. 
The medium is more persistent and farther from equilibrium for smaller $\alpha$, 
and the medium reduces to passive Brownian motion for very large $\alpha$.

\begin{figure}[!ht]
    \centering
    \includegraphics[scale=0.1]{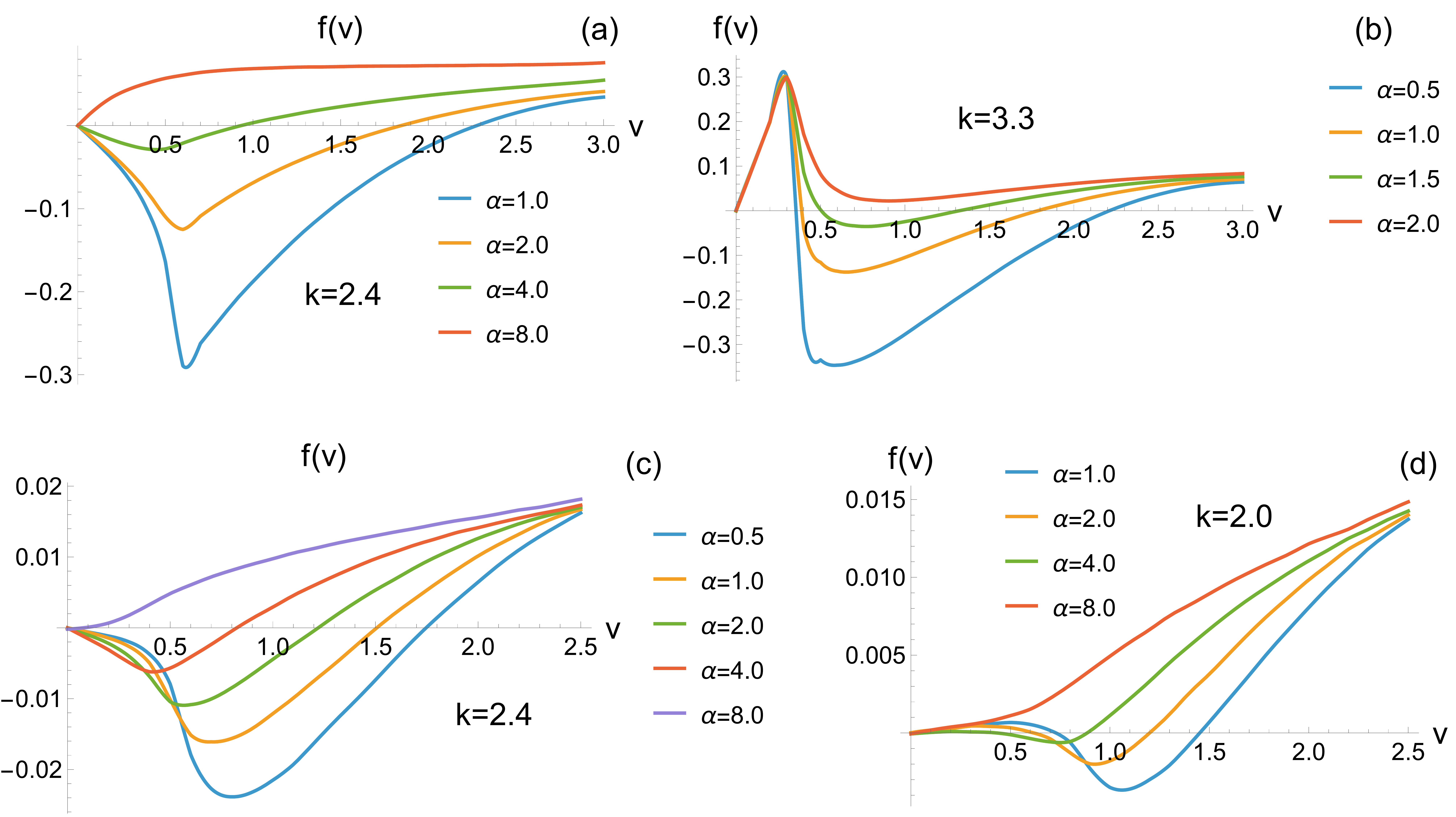}
    \caption{$f(v)$ for different $\alpha$: (a)(b) for 1D run-and-tumble medium and soft interaction, other parameters are $L=10$, $R=0.5$, $\mu=1$, $u=3$; (c)(d) for 2D active Brownian medium and Lennard-Jones interaction, other parameters are $L=10$, $R=0.5$, $\mu=1$, $u=3$.}\label{f1dalpha}
\end{figure}

%Negativity of $f(v)$ is possible only for small $\alpha$.
%We now further investigate how $f(v)$ changes with respect to $\alpha$. 
We plot $f(v)$ while fixing the interaction strength and changing $\alpha$, 
shown in Fig. \ref{f1dalpha}(a)(b) and Fig. \ref{f1dalpha}(c)(d) respectively for 1D and 2D cases. 
%The left panel and right panel represent 1D and 2D cases respectively.

For large $\alpha$, $f(v)$ is purely positive. 
As $\alpha$ decreases, we see the negativity of $f(v)$ appears. 
Depending on the interaction strength, the sign structure of $f(v)$ is either $(-,+)$ or $(+,-,+)$.
The former sign structure corresponds to active regime (R2a)/(A2a) while the latter sign structure corresponds to active regime (R2b)/(A2b) in 1D/2D.
%Correspondingly, the probe motion changes from passive regime to active regimes. 

If we continue to decrease the $\alpha$, the negative range of $f(v)$ grows monotonically. 
This indicates that for smaller $\alpha$, the active propulsion speed $v^*$ (satisfying $f(v^*)=0$, $f^\prime(v^*)>0$) of the probe is larger. 
However, $v^*$ can never exceed the propulsion speed of medium particles $u$ ($=3$ in Fig. \ref{f1dalpha}), even for $\alpha\rightarrow 0$. 
This is because if $v\geq u$, the medium particles that move in the same direction of probe cannot catch up with the probe and the negative friction mechanism ceases to operate.
Therefore, $v^*$ saturates at a value smaller than $u$ in the limit of  $\alpha\rightarrow0$. 

Meanwhile, we notice that changing $\alpha$ cannot lead to transitions between two sign structures $(-,+)$ and $(+,-,+)$.
%Therefore, the concrete active regime of the probe motion is not related to the medium persistence.
As we will see below, the concrete active regime of the probe motion is determined by the interaction strength. 

\subsection{Dependence on Interaction Strength}
The interaction strength plays an important role in the active transfer phenomenon. 
We consider 1D and 2D cases separately. 

For 1D case, 
we fix $\alpha=1.5$ (a relatively small value) and plot $f(v)$ as the interaction strength $k$ varies in Fig. \ref{f1dint}. 

The results align with the theoretical predictions in the End Matter: the profile of 
$f(v)$ for a highly persistent medium is determined by the relative magnitude between maximum interaction ($k$) and medium propulsion force ($\mu u$). 

When $k<\mu u$, 
the sign structure of $f(v)$ is $(-,+)$, shown in Fig. \ref{f1dint}(a), and the probe motion is in active regime (R2a).
In this case, the medium particles can pass through the probe easily as long as $v$ is not too large. 

For $k\gtrsim \mu u$, 
the system belongs to active regime (R2b).
As $k$ surpasses $\mu u$,
$f(v)$ suffers from a singular change from $(-,+)$ to $(+,-,+)$; see Fig. \ref{f1dint}(b). 
A positive range appear for small $v$. 
This is because the medium particles can no longer pass through the probe for small $v$.
As $k$ increases, that positive range grows.

For very large $k$, the negative range disappears completely, 
and the probe enters passive regime (R1).
In this case, the interaction is effectively hardcore, and it is very hard for medium particles to pass through the probe.
\begin{figure}[h]
    \centering
    \includegraphics[scale=0.1]{f1dint.pdf}
    \caption{$f(v)$ for different $k$ for 1D case: other parameters are $L=10$, $R=0.5$, $\mu=1$, $u=3$, $\alpha=1.5$. }\label{f1dint}
\end{figure}

The 2D case is more complicated, as 
the overall friction results from the competition of two contributions: colliding with hardcore (positive) or passing through the soft periphery (negative). 
While fixing $\alpha=1.5$, we plot $f(v)$ 
as $k$ (the maximum value of attraction in Lennard-Jones force) changes in Fig. \ref{f2dint}.

For very small $k$ (Fig. \ref{f2dint}(a)), 
the friction is purely positive, and the probe is in passive regime (A1). 
In this case, the interaction in soft periphery is too small and the collision effect dominates. 

As $k$ increases (Fig. \ref{f2dint}(b)), the sign structure of $f(v)$ becomes $(+,-,+)$, and the probe motion is in active regime (A2b).
In this case, two contributions are comparable, and friction is negative only in a moderate range of velocity $(v^\dagger,v^*)$.
As $k$ increases, the negative range grows.

For larger $k$ but still $k<\mu u$ (Fig. \ref{f2dint}(c)(d)), 
the sign structure is $(-,+)$, and the probe motion changes to active regime (A2a). 
%The negativity becomes prominent as $k$ increases. 

For $k\gtrsim \mu u$ (Fig. \ref{f2dint}(e)), the probe motion belongs to active regime (A2b) again. 
As $k$ surpasses $\mu u$, a singular transition from $(-,+)$ to $(+,-,+)$ occurs, similar to 1D. 
%Recall that $k$ represents the maximum attractive interaction. 
In this case, medium particles are trapped by the probe if $v$ is small, similar to the case of being stuck in 1D, resulting in a positive range of $f(v)$ for small $v$. 

For very large $k$ (Fig. \ref{f2dint}(f)), 
the negativity of $f(v)$ disappears.
We have passive regime (A1) for the probe motion. 
In this case, most medium particles are trapped by the probe.
\begin{figure}[htp]
    \centering
    \includegraphics[scale=0.1]{f2dint.pdf}
    \caption{$f(v)$ for different $k$ for 2D case: other parameters are $L=10$, $R=0.5$, $\mu=1$, $u=3$, $\alpha=1.5$. }\label{f2dint}
\end{figure}

In summary, 
we find that the interaction strength crucially determines the concrete active regimes. 
For a high-persistent medium, as $k$ changes, 
we can observe all three regimes (R1)(R2a)(R2b) or (A1)(A2a)(A2b) in both 1D and 2D. 
In 1D, it can be deduced by whether medium particles are stuck by the probe. 
In 2D, it is more complicated since the friction results from a competition between two contributions.

\section{Simulation details}
Two kinds of numerical simulations are involved in our study. 
One is to calculate the landscapes of $f(v)$, $B(v)$, and $G(v)$ appearing in the reduced dynamics by evaluating the expectation values in the fixed-$v$ dynamics. 
The other is to simulate the original composite dynamics, whose results are used to test the validity of our reduced dynamics. 

\subsection{Calculation of landscapes of velocity-dependent friction and noise}
In the reduced dynamics, the first-order nonlinear friction $f(v)$, its second-order correction $G(v)$, and the noise intensity $B(v)$ are expressed as expectation values in the fixed-$v$ dynamics of a single medium particle, given by Eq.~\eqref{exp1d}~\eqref{exp2d}. 
Here, we roughly explain how we obtain the landscapes of $f(v)$, $G(v)$, and $B(v)$ through simple numerical calculations. 

% In the following, we only explain how the simulation works for 1D. 
% There is no essential difference in 2D. 
The dynamics under investigation is given by Eq.~\eqref{medium_dynamics} with $\vec v$ held fixed. 
We use a second-order stochastic Runge-Kutta method \cite{Honeycutt1992} to simulate it. 
In one small time step $\Delta t$, we update $r(t+\Delta t)$.

In the calculation of $f(v)$ and $B(v)$, we run the dynamics for $N$ independent replicas of medium particle. 
We start from an arbitrary distribution and let the system evolve for time period $T_\text{relax}$ to arrive at the stationary state.
Then, the program is run for time period $T_\text{record}$, and the interaction force $F(r(t))$ at each time is recorded. 
$f(v)$ (per medium particle) is obtained by averaging $F(r)$ over both the time steps during $T_\text{record}$ and the medium particle replicas. 
When evaluating $B(v)$, the time integral over $s$ in Eq.~\eqref{exp1d} is cutoff at $T_\text{correlation}$, which is chosen to be the time when the correlation function gets vanishingly small. 
The integrand $\ev{F(s);F(0)}$ is also obtained by averaging over both the time steps and the replicas.

The calculation of $G(v)$ is a bit tricky. 
We recall the expression of $G(v)$, 
\begin{equation}\label{Grepeat}
    G(v)=N\int_0^\infty \ud s\ev{\delta F(r(s))F(r(0))\partial_v\log\rho_v(r(0))}_v ,
\end{equation}
%The stationary distribution function of a medium particle $\rho_v(r)$ in the fixed-$v$ dynamics is involved. 
with $\delta F(r)=F(r)-\ev{F(r)}_v$. 
We usually do not know the expression of $\rho_v(r)$ since it is a stationary distribution of a nonequilibrium dynamics (fixed-$v$ dynamics).

To overcome this problem, we first notice the following relation
\begin{equation}\label{transition_probability}
\begin{aligned}
    &\ev{\delta F(r(s))F(r(0))\partial_v\log\rho_v(r(0))}_v\\
    =&\lim_{\ud v\rightarrow0}\int\ud r(0)\int\ud r(s)\,\delta F(r(s))F(r(0))\frac{1}{\ud v}\left[\rho_{v+\ud v/2}(r(0))-\rho_{v-\ud v/2}(r(0))\right]P_v(r(s)|r(0)),
    \end{aligned}
\end{equation}
where $P_v(r(s)|r(0))$ is the transition probability from $r(0)$ to $r(s)$ in the fixed-$v$ dynamics with probe velocity fixed to $v$, and $\rho_{v\pm\ud v/2}$ is the stationary distribution in the fixed-$v$ dynamics with probe velocity fixed to $v\pm \ud v/2$.

Now we introduce a time-dependent quench dynamics for a single medium particle. 
In Eq.~\eqref{medium_dynamics}, we put $v(t) = v_- $ for $t\in(-\infty,0)$ and $v(t)=v_+$ for $t\in(0,+\infty)$. 
That is, the probe velocity keeps $v=v_{-}$ from the infinite past until time $t=0$; at $t=0$, the probe velocity is suddenly and forever changed to $v_+$. 
As a result, the medium has distribution $\rho_{v_-}$ at $t=0$, and then undergoes a relaxation process with fixed $v_+$.  
We denote  $\ev{ F(q,\eta(s))}_{v_-\rightarrow v_+}$  for the average of $F$ in this time-dependent dynamics.
By using \eqref{Grepeat} and \eqref{transition_probability}, we find $G(v)$ can be rewritten as 
\begin{equation}\label{Gv_cal}
    G(v)=N\lim_{\ud v\rightarrow0}\frac1{\ud v}\int_0^\infty\ud s\left[\ev{\delta F(r(s))F(r(0))}_{v+\ud v/2\rightarrow v}-\ev{\delta F(r(s))F(r(0))}_{v-\ud v/2\rightarrow v}\right].
\end{equation}
In the above expression, $G(v)$ is related to correlation functions in two time-dependent dynamics, and the nonequilibrium distribution $\rho_v(r)$ does not appear anymore. 

In the simulation, we set a fixed value $\Delta v$ to replace the limit of $\ud v$.
We take two groups of replicas of medium particle, each containing $N_G$ particles. 
We evolve the dynamics for time $T_\text{relax}$ under $v+\Delta v/2$ and $v-\Delta v/2$, respectively for each group. 
Then, we change the parameter to be $v$ suddenly for both groups and perform the second evolution for time $T_\text{relax}$. 
The relaxation time $T_\text{relax}$ is chosen to make sure the relaxation. 
In the evolution after the sudden quench, we record the interaction force $F(r)$ for each step and for each medium particle replica. 
For each group, the integral over $s$ in Eq.~\eqref{Gv_cal} is evaluated after taking the ensemble average of $\delta F(s)F(0)$ over replicas, and the upper limit of the integral is cut off at $T_\text{relax}$.
Finally, $G(v)$ is calculated from a finite difference between the integral of two groups. 
Since $G(v)$ is calculated from a differential form, 
getting a precise results of $G(v)$ needs more replicas $N_G$ than that used in calculating $f(v)$ and $B(v)$.

\subsection{Simulation of composite dynamics}
We also simulate the dynamics of the composite system directly. 
We aim to obtain the stationary distribution of the probe velocity $\rho^\text{st}(v)$. 
Meanwhile, in the case that velocity distribution has discrete peaks, for example, regimes (R2a) (R2b) in 1D and regime (A2b) in 2D described in the main text, we aim to obtain transition rates between those stable velocities. 
These two results (stationary distribution and transition rates) are used to verify our reduced dynamics. 

We choose the medium particle number $N$ to be a small number ($N\sim5$ while $L=10$) to ensure the dilute limit. 
We apply a second-order stochastic Runge-Kutta method to simulate the composite dynamics. 
The time step $\Delta t$ is chosen to be much smaller than the typical time-scale of the medium particles, and the total simulation time is chosen to be large enough so that transitions between different velocities have happened for many times. 

After the composite system relaxes to a steady state, we save the probe's velocity every time interval $N_\text{save\_rate}\Delta t$. 
From the data of the probe's velocity, we reconstruct the real stationary distribution of velocity through a kernel density estimation (KDE) method.

Besides, in the simulation, whenever a transition from a stable velocity to another stable velocity happens, we save the current time as the (absolute) transition time. 
The difference of two consecutive transition times yields a passage time between two stable velocities. 
The mean first passage time is obtained from the average of all passage times observed in the simulation, and its inverse determines the transition rate observed from the simulation. 

% we save another set of data 
% The transition time is determined as follows. 
% Suppose $v_1$ and $v_2$ are two stable velocities, and $v_0$ is the barrier point between them (where the stationary distribution is minimized).
% If the probe velocity, starting from a velocity near $v_1$, overcomes the barrier and passes $(v_2+v_0)/2$ at some time step, then we save the current time and identify it as the (absolute) transition time from $v_1$ to $v_2$. 

% The stationary distribution from the reduced dynamics is given by
% \begin{equation}
%     \rho(v)\propto\exp(-\phi(v))
% \end{equation}
% with 
% \begin{equation}
%     \phi=\int^v\ud w \frac{M f(v)+G(v)}{B(v)}
% \end{equation}

% In 2D, the stationary distribution of the probe velocity 
% \begin{equation}
%     \rho(v_x,v_y)\propto\exp(-\phi(v))
% \end{equation}
% with an effective 1D potential 
% \begin{equation}
%     \phi = \int^v\ud w \frac{M f(v) +G(v)+[B_\parallel(v)-B_\perp(v)]/v}{B_\parallel(v)}
% \end{equation}
% We also find that, in the total nonlinear friction up to the second order, $g(v)=f(v)+G(v)/M$, 
% the second term cannot be neglected in a quantitative comparison, even if in large $M$ limit, it is relatively small compared to $f(v)$. 

\bibliography{transmit}